\documentclass[aps,floats]{revtex4}
\usepackage{amsmath,amssymb}
\usepackage{graphicx,epsfig}
\usepackage[greek,english]{babel}

\begin{document}
\bibliographystyle {plain}

\def\oppropto{\mathop{\propto}} 
\def\opsimeq{\mathop{\simeq}}
\def\opoverderline{\mathop{\overline}}
\def\operarrow{\mathop{\longrightarrow}}
\def\opsim{\mathop{\sim}}

\def\fig#1#2{\includegraphics[height=#1]{#2}}
\def\figx#1#2{\includegraphics[width=#1]{#2}}


\title{ Strong Disorder Real-Space Renormalization \\
for the Many-Body-Localized phase of random Majorana models   } 


\author{ C\'ecile Monthus }
 \affiliation{Institut de Physique Th\'{e}orique, 
Universit\'e Paris Saclay, CNRS, CEA,
91191 Gif-sur-Yvette, France}

\begin{abstract}
For the Many-Body-Localized phase of random Majorana models, a general Strong Disorder Real-Space Renormalization procedure known as RSRG-X [D. Pekker, G. Refael, E. Altman, E. Demler and V. Oganesyan, Phys. Rev. X 4, 011052 (2014)] is described to produce the whole set of excited states, via the iterative construction of the Local Integrals of Motion (LIOMs). The RG rules are then explicitly derived for arbitrary quadratic Hamiltonians (free-fermions models) and for the Kitaev chain with local interactions involving even numbers of consecutive Majorana fermions. The emphasis is put on the advantages of the Majorana language over the usual quantum spin language to formulate unified RSRG-X rules.

\end{abstract}

\maketitle

\section{ Introduction }

Strong Disorder Renormalization procedures introduced for ground-states of random quantum models 
by Ma-Dasgupta-Hu \cite{ma-dasgupta1,ma-dasgupta2} and Daniel Fisher \cite{fisher_AF,fisher,fisherreview} (see the review \cite{strong_review} and references therein) 
are usually formulated in terms of quantum spins. Although one can indeed argue that the language of quantum spins $S=1/2$ or q-bits
is the most natural framework for quantum models or quantum information, another appealing point of view is 
that it is much more advantageous to use instead the language of Majorana fermions
in order to reveal the true underlying structure of the model, 
that could be otherwise somewhat hidden in the spin formulation (see for instance the two recent works \cite{moessner,10phases} where the Majorana language
is instrumental to classify possible phases).

In the present paper, the goal is thus to formulate Strong Disorder Renormalization rules for generic random Majorana models.
Besides the construction of the ground-state mentioned above, the Strong Disorder Renormalization approach has been recently extended 
to construct the whole set of excited eigenstates via the RSRG-X procedure 
 \cite{rsrgx,rsrgx_moore,vasseur_rsrgx,yang_rsrgx,rsrgx_bifurcation,rsrgx_bifurcation_xyz},
or to obtain the effective dynamics via the RSRG-t procedure \cite{vosk_dyn1,vosk_dyn2}.
These two closely related procedures \cite{c_rsrgt} actually identify iteratively  the Local Integrals of Motion called LIOMs
\cite{emergent_swingle,emergent_serbyn,emergent_huse,emergent_ent,imbrie,serbyn_quench,liom_lightcone,emergent_vidal,emergent_ros,
emergent_rademaker,serbyn_powerlawent,liom_obrien,c_emergent,bhatt_liom,ros_remanent,wortis_liom,c_liom,marco_flow,maj_liom,counting_liom}
that are known to characterize the Many-Body-Localized phase existing in some isolated random quantum interacting models (see the many recent reviews \cite{revue_huse,revue_altman,revue_vasseur,revue_imbrie,revue_rademaker,review_mblergo,review_prelovsek,review_rare,review_toulouse} and references therein).

The paper is organized as follows.
In section \ref{sec_models}, the notations for general random Majorana Models with parity-interactions are introduced.
In section \ref{sec_rsrgx}, the general RSRG-X procedure is described with the simplest example of the random Kitaev chain.
In section \ref{sec_free}, the RSRG-X rules are given for arbitrary quadratic Hamiltonians (free-fermions). 
In section \ref{sec_consecutive}, the RSRG-X rules are derived for the random Majorana chain with local interactions
involving only consecutive Majorana operators.
The conclusions are summarized in section \ref{sec_conclusion}.
The Appendix \ref{sec_app} contains a short reminder of
the dictionary between Majorana fermions, Dirac fermions and quantum spin chains.

\section{ Notations for random Majorana Models with parity-interactions }

\label{sec_models}

\subsection{ Majorana operators }

In the present paper, we wish to study models defined in terms
of $2N$ Majorana operators $\gamma_j$ with $j=1,..,2N$ (see Appendix \ref{sec_app} for 
the dictionary between Majorana fermions, Dirac fermions and quantum spin chains.).
These Majorana operators are hermitian
\begin{eqnarray}
\gamma_j^{\dagger}=\gamma_j
\label{hermi}
\end{eqnarray}
 square to unity 
\begin{eqnarray}
\gamma_j^2=1 
\label{squareunity}
\end{eqnarray}
and anti-commute with each other
\begin{eqnarray}
\{ \gamma_j , \gamma_l \} \equiv \gamma_j \gamma_l + \gamma_l \gamma_j && = 0 \ \ \ \ \ \  { \rm for } \ \ \   j \ne l
\label{anticomm}
\end{eqnarray}
So the first advantage of the Majorana formulation over Dirac fermions or quantum spins
is clearly the symmetric role played by the $2N$ Majorana operators instead of the creation and annihilation operators for  the Dirac fermions,
or the three Pauli matrices for quantum spins (see Appendix \ref{sec_app}).
One thus expects that the Majorana language is more appropriate to formulate unified renormalization rules.

\subsection{ Parity operators }

It is convenient to associate to any even number $(2k)$ with $k=1,2,..N$ of Majorana operators labelled by $1 \leq j_1<j_2<..<j_{2k} \leq 2N$
the parity operator
\begin{eqnarray}
P^{(2k)}_{j_1,j_2,..,j_{2k}} \equiv    i^k \gamma_{j_1} \gamma_{j_2}\gamma_{j_3} \gamma_{j_4}  ... \gamma_{j_{2k-1}} \gamma_{j_{2k}}
\label{parityj}
\end{eqnarray}
For $k=1$ and $k=2$, they represent the usual interactions between two and four Majorana operators respectively
\begin{eqnarray}
P^{(2)}_{j_1,j_2}  && = i \gamma_{j_1} \gamma_{j_2} 
\nonumber \\
P^{(4)}_{j_1,j_2,j_3,j_4} &&= - \gamma_{j_1} \gamma_{j_2} \gamma_{j_3} \gamma_{j_4}
\label{parityj24}
\end{eqnarray}
while for $k=N$, the only possibility is $j_q=q$ leads to the standard total parity of the whole system
\begin{eqnarray}
P^{tot} \equiv  P^{(2N)}_{1,2,..,2N-1,2N}  = i^N \gamma_1 \gamma_2 \gamma_{3} \gamma_{4}   ... \gamma_{2N-1} \gamma_{2N}
\label{paritytotal}
\end{eqnarray}
The parity operators of Eq. \ref{parityj} are hermitian
\begin{eqnarray}
(P^{(2k)}_{j_1,j_2,..,j_{2k}}  )^{\dagger} = P_{j_1,j_2,..,j_{2k}}  
\label{parityjh}
\end{eqnarray}
square to unity
\begin{eqnarray}
(P^{(2k)}_{j_1,j_2,..,j_{2k}} )^2    = 1
\label{parityjsquare}
\end{eqnarray}
and they commute or anti-commute 
\begin{eqnarray}
P^{(2k)}_{j_1,j_2,..,j_{2k}} P^{(2q)}_{l_1,l_2,..,l_{2q}} = (-1)^{p_c} P_{l_1,l_2,..,l_{2q}} P_{j_1,j_2,..,j_{2k}} 
\label{parity2prod}
\end{eqnarray}
depending on the parity $(-1)^{p_c}$ of the number $p_c$ of common Majorana operators between the two sets $\{j_1,,,,j_{2k}\}$
and $\{l_1,,,,l_{2q}\}$.

\subsection{ General Hamiltonian commuting with the total parity  }

The most general hermitian Hamiltonian commuting with the total parity $P^{tot}$ of Eq. \ref{paritytotal}
can be expanded into all the parity operators of Eq. \ref{parityj}
\begin{eqnarray}
{\cal H} && = \sum_{k=1}^N  {\cal H}^{(2k)}
\nonumber \\
 {\cal H}^{(2k)} && = \sum_{1 \leq j_1<j_2<..<j_{2k-1}< j_{2k} \leq 2N} K^{(2k)}_{j_1,j_2,..j_{2k}}  P^{(2k)}_{j_1,j_2,..,j_{2k}}
\label{HKparity}
\end{eqnarray}
where $K^{(2k)}_{j_1....j_{2k}} $ are the real couplings defining the model.

For instance, $ {\cal H}^{(2)}  $ corresponds to the most general quadratic Hamiltonian
\begin{eqnarray}
 {\cal H}^{(2)} && =  \sum_{1 \leq j_1<j_2\leq 2N}   K^{(2)}_{j_1 j_{2}}   P^{(2)}_{j_1 j_{2}}  
= i \sum_{1 \leq j_1<j_2\leq 2N}   K^{(2)}_{j_1 j_{2}}  \gamma_{j_1}\gamma_{j_2} 
\label{h2}
\end{eqnarray}
while $ {\cal H}^{(4)}  $ contains all the possible four-Majorana-interactions
\begin{eqnarray}
 {\cal H}^{(4)} && = \sum_{1 \leq j_1<j_2<j_3< j_{4} \leq 2N} K^{(4)}_{j_1,j_2,j_3,j_{4}}  P^{(4)}_{j_1,j_2,j_3,j_{4}}  
= - \sum_{1 \leq j_1<j_2<j_3< j_{4} \leq 2N} K^{(4)}_{j_1,j_2,j_3,j_{4}}  \gamma_{j_1} \gamma_{j_2} \gamma_{j_3} \gamma_{j_4}
\label{h4}
\end{eqnarray}

Before specializing to specific models, it is useful to define first a general RSRG-X procedure for the generic Hamiltonian of Eq. \ref{HKparity},
as described in the following section.

\section{ General RSRG-X procedure for random Majorana models}

\label{sec_rsrgx}

In this section, we consider the generic Majorana Hamiltonian of Eq. \ref{HKparity}
with random coupling constants $ K^{(2k)}_{j_1....j_{2k}}$, and we describe the RSRG-X procedure
based on the decimation of the strongest two-Majorana-coupling.

\subsection{ Strongest two-Majorana-coupling   }

Let us choose the biggest two-Majorana-coupling in absolute value $\vert K^{(2)} _{n m} \vert $ with $1 \leq n<n\leq 2N $ 
\begin{eqnarray}
\vert K^{(2)} _{n m} \vert && = \max_{1 \leq j_1<j_2\leq 2N}  \vert  K^{(2)} _{j_1 j_{2}}  \vert
\label{k2max}
\end{eqnarray}

The corresponding elementary two-Majorana Hamiltonian involves only the parity $ P^{(2)}_{nm}$
\begin{eqnarray}
h_{nm} && =   K^{(2)} _{nm} P^{(2)} _{nm} =  i K^{(2)} _{nm}  \gamma_{n}\gamma_{m} 
\label{h0}
\end{eqnarray}
so the two eigenvalues labelled by $\epsilon=\pm 1$
\begin{eqnarray}
e_{nm}^{\epsilon} && =  \epsilon K^{(2)} _{nm}
\label{e0}
\end{eqnarray}
are associated to the two orthogonal projectors 
 \begin{eqnarray}
\pi_{nm}^{\epsilon} && = \frac{1 +\epsilon P^{(2)} _{nm}  }{2}  = \frac{1 + i \epsilon \gamma_{n}\gamma_{m}    }{2}
\label{projnm}
\end{eqnarray}

\subsection{ Perturbation theory in the other couplings   }

The projection of the full Hamiltonian on the two energy branches labelled by $\epsilon=\pm 1$ (Eqs \ref{e0}) 
reads at second-order perturbation theory 
in all the other couplings 
\begin{eqnarray}
{\cal H}^{eff} && = {\cal H}^{+}_{nm}  + {\cal H}^{-}_{nm} +O\left(  \frac{1}{K_{nm}^2} \right)
\nonumber \\
{\cal H}^{\epsilon}_{nm} && =\pi_{nm}^{\epsilon} {\cal H}  \pi_{nm}^{\epsilon} + 
 \frac{( \pi_{nm}^{\epsilon} {\cal H} \pi_{nm}^{-\epsilon}) ( \pi_{nm}^{-\epsilon} {\cal H}  \pi_{nm}^{\epsilon}) }{ 2 \epsilon K^{(2)}_{nm}   } 
\label{veff}
\end{eqnarray}

To evaluate how the Hamiltonian ${\cal H}$ behaves between two equal $\epsilon=\epsilon'$ or opposite $\epsilon =- \epsilon'$ projectors of Eq. \ref{projnm}, it is useful to decompose ${\cal H}$ into the four terms 
 \begin{eqnarray}
 {\cal H} 
= H_{nm}^{00}  + i \gamma_{n}  H_{nm}^{10}  + i \gamma_{m} H_{nm}^{01} + i \gamma_{n}\gamma_{m}  H_{nm}^{11}
\label{decomponm}
\end{eqnarray}
where the $H^{\alpha_n \alpha_m}_{n m}$ involve only the other Majorana fermions $(\gamma_j )  $ with $j \ne (n,m)$.
In particular, $ H_{nm}^{00}   $ and $ H_{nm}^{11}  $ contain an even number of these other Majorana operators,
while $ H_{nm}^{10}  $ and $  H_{nm}^{01} $ contain an odd number of these other Majorana operators.
As a consequence, the part ${\cal H}_{nm}^{comm}   $ of ${\cal H}$  that commutes with the parity $P^{(2)}_{nm}= i \gamma_{n}\gamma_{m}  $ reads
 \begin{eqnarray}
{\cal H}_{nm}^{comm} = H_{nm}^{00}  + i \gamma_{n}\gamma_{m}  H_{nm}^{11}
\label{ocom}
\end{eqnarray}
while the contribution $ {\cal H}_{nm}^{anti} $ of ${\cal H}$ that anticommutes with the parity $P^{(2)}_{nm}= i \gamma_{n}\gamma_{m}$ is
 \begin{eqnarray}
 {\cal H}_{nm}^{anti}  = i \gamma_{n}  H_{nm}^{10}  + i \gamma_{m} H_{nm}^{01} 
\label{hanti}
\end{eqnarray}

Between two identical projectors $\epsilon=\epsilon'$,
only the commuting part survives and gives the contribution
 \begin{eqnarray}
 \pi_{nm}^{\epsilon}  {\cal H} \pi_{nm}^{\epsilon} = \pi_{nm}^{\epsilon}  {\cal H}^{comm} \pi_{nm}^{\epsilon}  && = H_{nm}^{00}  + \epsilon H_{nm}^{11}
\label{projrules}
\end{eqnarray}
Between two orthogonal projectors, only the anticommuting part survives and yields
\begin{eqnarray}
\pi_{nm}^{\epsilon} {\cal H}    \pi_{nm}^{-\epsilon}  = \pi_{nm}^{\epsilon} {\cal H}_{nm}^{anti}    \pi_{nm}^{-\epsilon} = \pi_{nm}^{\epsilon} H_{nm}^{anti}  = H_{nm}^{anti}    \pi_{nm}^{-\epsilon}
\label{projorthorules}
\end{eqnarray}
so that the the numerator of Eq. \ref{veff} becomes
\begin{eqnarray}
( \pi_{nm}^{\epsilon} {\cal H} \pi_{nm}^{-\epsilon}) ( \pi_{nm}^{-\epsilon} {\cal H} \pi_{nm}^{\epsilon}) = 
\pi_{nm}^{\epsilon} ({\cal H}_{nm}^{anti})^2 \pi_{nm}^{\epsilon}
\label{oanti2}
\end{eqnarray}
Since $ H_{nm}^{01} $ and $ H_{nm}^{10} $ contain an odd number of the other Majorana operators $(\gamma_j )  $ with $j \ne (n,m)$, one obtains
that the square of Eq. \ref{hanti} reads
 \begin{eqnarray}
 (H_{nm}^{anti})^2 && = - \left( \gamma_{n}  H_{nm}^{10}  + \gamma_{m} H_{nm}^{01} \right)^2
=  (  H_{nm}^{10} )^2 + (  H_{nm}^{01} )^2 - \gamma_n \gamma_m [H_{nm}^{01}, H_{nm}^{10}   ]
\label{decompanti2}
\end{eqnarray}
so that its projection reads
\begin{eqnarray} 
\pi_{nm}^{\epsilon}  (H_{nm}^{anti})^2 \pi_{nm}^{\epsilon} = 
   (  H_{nm}^{10} )^2 + (  H_{nm}^{01} )^2 +i \epsilon [H_{nm}^{01}  , H_{nm}^{10} ]
\label{projoanti2}
\end{eqnarray}

Putting everything together, one obtains that the effective Hamitonian 
of Eq \ref{veff} for the remaining Majorana operators reads
\begin{eqnarray}
H^{\epsilon}_{nm} && = H_{nm}^{00}  + \epsilon H_{nm}^{11}
+ \epsilon  \frac{   (  H_{nm}^{10} )^2 + (  H_{nm}^{01} )^2  }{ 2 K^{(2)}_{nm}   } 
+  \frac{  i  [H_{nm}^{01}  , H_{nm}^{10} ]  }{ 2  K^{(2)}_{nm}   } 
\label{veffres}
\end{eqnarray}
in terms of the decomposition of Eq. \ref{decomponm}.
To see how this procedure works in practice, let us now describe the simplest possible case.

\subsection{ Simplest application : the random Kitaev chain }

As recalled in Appendix \ref{sec_app}, the Kitaev chain \cite{kitaevchain} with random nearest-neighbor-two-Majorana couplings $K^{(2)}_{j,j+1} $
 \begin{eqnarray}
H^{Kitaev} = i \sum_{j=1}^{2N-1} K^{(2)}_{j,j+1} \gamma_j \gamma_{j+1}
\label{kitaev}
\end{eqnarray}
corresponds to the Random Transverse Field Ising Chain (RTFIC) of Eq. \ref{kitaevspins}.
Since the RTFIC is one of the basic model where the Strong Disorder RG approach has been developed \cite{fisher},
it is useful to mention how the RSRG-X procedure described above works for the random Kitaev chain of Eq. \ref{kitaev}.

One chooses the biggest coupling in absolute value (Eq. \ref{k2max})
\begin{eqnarray}
\vert K^{(2)} _{n,n+1} \vert && = \max_{1 \leq j \leq 2N-1}  \vert  K^{(2)} _{j, j+1}  \vert
\label{k2maxkitaev}
\end{eqnarray}
and
one computes the corresponding decomposition of Eq. \ref{decomponm}
\begin{eqnarray}
 {\cal H} 
&& = H_{n,n+1}^{00}  + i \gamma_{n}  H_{n,n+1}^{10}  + i \gamma_{n+1} H_{n,n+1}^{01} + i \gamma_{n}\gamma_{n+1}  H_{n,n+1}^{11}
\nonumber \\
H_{n,n+1}^{00}  && =i \sum_{j=1}^{n-2} K^{(2)}_{j,j+1} \gamma_j \gamma_{j+1} + i \sum_{j=n+2}^{2N-1} K^{(2)}_{j,j+1} \gamma_j \gamma_{j+1}
\nonumber \\
H_{n,n+1}^{10}  && =- K^{(2)}_{n-1,n} \gamma_{n-1} 
\nonumber \\
H_{n,n+1}^{01}  && =K^{(2)}_{n+1,n+2}  \gamma_{n+2}
\nonumber \\
H_{n,n+1}^{11}  && = K^{(2)}_{n,n+1} 
\label{decomponmkitaev}
\end{eqnarray}
in order to obtain the effective Hamiltonian via Eq. \ref{veffres}
\begin{eqnarray}
H^{\epsilon}_{n,n+1} && = H_{nm}^{00} 
 + \epsilon \left(  K^{(2)}_{n,n+1} 
+   \frac{   ( K^{(2)}_{n-1,n}  )^2 + ( K^{(2)}_{n+1,n+2}  )^2  }{ 2 K^{(2)}_{n,n+1}   } \right)
+ i \frac{  K^{(2)}_{n-1,n}    K^{(2)}_{n+1,n+2}    }{   K^{(2)}_{n,n+1}    } \gamma_{n-1} \gamma_{n+2}
\label{veffreskitaev}
\end{eqnarray}
So besides the first term $H_{nm}^{00}  $ representing the part of the chain that is left unchanged by the decimation of the pair $(\gamma_{n} ,\gamma_{n+1})$
and the second term proportional to $\epsilon$ representing the direct energy contribution of the decimation, 
the third term means that the Majorana operators $\gamma_{n-1} $ and $\gamma_{n+2}$ that become nearest-neighbor after the decimation
are now coupled by the renormalized coupling
\begin{eqnarray}
K^{r(2)}_{n-1,n+2} =  \frac{  K^{(2)}_{n-1,n}    K^{(2)}_{n+1,n+2}    }{   K^{(2)}_{n,n+1}    }
\label{kr2kitaev}
\end{eqnarray}
that is independent of the energy branch $\epsilon=\pm 1$ chosen for the decimation.
This independence is of course not surprising, since it is a direct consequence of the notion of 'free fermions',
but it is nevertheless important to stress here the difference with the RSRG-X rules formulated in the spin language,
where the choice $\epsilon=\pm 1$ of the energy branch explicitly appear in the renormalization of the couplings \cite{rsrgx}.

In the remainder of the paper, we analyze two different generalizations of this random Kitaev chain.
We first describe how the RSRG-X procedure works for arbitrary quadratic Hamiltonians in section \ref{sec_free}.
We then consider the random Kitaev chain in the presence of local interactions involving even numbers of consecutive Majorana fermions
in section \ref{sec_consecutive}.

\section{ Application to arbitrary quadratic Hamiltonians }

\label{sec_free}

In this section, the RSRG-X procedure described in the previous section
is applied to any random quadratic Hamiltonians (free-fermions).

\subsection{ Decomposition of Eq. \ref{decomponm}  }

When the Hamiltonian contains only pair-interaction between Majorana operators (only $k=1$ in Eq. \ref{HKparity})
\begin{eqnarray}
{\cal H} = {\cal H}^{(2)} && =   \sum_{1 \leq j_1<j_2 \leq 2N} K^{(2)}_{j_1,j_2}  P^{(2)}_{j_1,j_2} = i \sum_{1 \leq j_1<j_2 \leq 2N} K^{(2)}_{j_1,j_2}  \gamma_{j_1} \gamma_{j_2} 
\label{H2}
\end{eqnarray}
the decomposition of Eq. \ref{decomponm} with respect to the pair $(\gamma_n \gamma_m)$ reads
 \begin{eqnarray}
 H_{nm}^{00}  && = \sum_{1 \leq j_1 <  j_2\leq 2N, j_1 \ne (n,m), j_2 \ne (n,m)} i K^{(2)}_{j_1j_2} \gamma_{j_1} \gamma_{j_2} 
\nonumber \\
 H_{nm}^{10}  &&  = \sum_{1 \leq j \leq n-1}  (-K^{(2)}_{jn} )\gamma_j   +    \sum_{n+1 \leq j \leq 2N, j \ne m}  K^{(2)}_{nj} \gamma_j     = \sum_{j \ne (n,m) }  K^{(2)}_{nj} \gamma_j
\nonumber \\
 H_{nm}^{01}  && = \sum_{1 \leq j \leq m-1,j \ne n}  (-K^{(2)}_{jm} ) \gamma_j   +    \sum_{m+1 \leq j \leq 2N}  K^{(2)}_{mj} \gamma_j    = \sum_{j \ne (n,m)}  K^{(2)}_{mj} \gamma_j
\nonumber \\
 H_{nm}^{11}  && =   K_{nm}  
\label{decompah2}
\end{eqnarray}
where we have introduced the notation for $j_2>j_1$
 \begin{eqnarray}
K^{(2)}_{j_2,j_1}= && = - K^{(2)}_{j_1,j_2}
\label{antisymk}
\end{eqnarray}

Since $ H_{nm}^{10}  $ and $ H_{nm}^{01}  $ are linear in the other Majorana operators,
their squares are constants
 \begin{eqnarray}
( H_{nm}^{10}  )^2 && =  \sum_{j \ne (n,m)}  (K^{(2)}_{nj})^2 
\nonumber \\
( H_{nm}^{01})^2  && = \sum_{j \ne (n,m)}  (K^{(2)}_{mj})^2
\label{twosquares}
\end{eqnarray}
while their commutator is quadratic
 \begin{eqnarray}
 [ H_{nm}^{01} , H_{nm}^{10} ]  && =  \sum_{j_1 \ne (n,m)}  K^{(2)}_{nj_1 }\sum_{j_2 \ne (n,m)}  K^{(2)}_{mj_2} [ \gamma_{j_1} , \gamma_{j_2} ]
\nonumber \\
&& =   \sum_{ j_1 <  j_2, j_1 \ne (n,m), j_2 \ne (n,m)} \left(   K^{(2)}_{nj_1} K^{(2)}_{m j_2} - K^{(2)}_{nj_2} K^{(2)}_{m j_1}\right) 2 \gamma_{j_1} \gamma_{j_2} 
\label{commutator}
\end{eqnarray}

\subsection{ RSRG-X rules  }

Putting everything together, Eq \ref{veffres} becomes
\begin{eqnarray}
H^{\epsilon}_{nm} && =  H_{nm}^{00}  +\epsilon \left(   K^{(2)}_{nm}  + \sum_{j \ne (n,m) } \frac{    (K^{(2)}_{nj})^2 +  (K^{(2)}_{mj})^2   }{ 2  K^{(2)}_{nm}   }  \right)
+  i \sum_{ j_1 <  j_2, j_1 \ne (n,m), j_2 \ne (n,m)}  K^{R(2)}_{j_1j_2} \gamma_{j_1} \gamma_{j_2} 
\label{veffres2}
\end{eqnarray}
with the renormalized couplings between the remaining Majorana operators
\begin{eqnarray}
K^{R(2)}_{j_1j_2} =  K^{(2)}_{j_1j_2}  + \frac{\left(   K^{(2)}_{nj_1} K^{(2)}_{m j_2} - K^{(2)}_{nj_2} K^{(2)}_{m j_1}\right)  }{   K^{(2)}_{nm}   } 
\label{kr2}
\end{eqnarray}
These RSRG-X rules are thus closed for any quadratic Hamiltonian, and represent a direct generalization of the rule discussed above for the Kitaev chain in Eq. \ref{kr2kitaev}.
Again, the choice of the energy branch $\epsilon=\pm 1$ appears only in the constant energy contribution of the decimation (second term of Eq. \ref{veffres2})
but not in the renormalized couplings of Eq. \ref{kr2} as a consequence of the notion of 'free-fermions'.

\section{ Application to the Majorana chain with consecutive-parity-interactions }

\label{sec_consecutive}

After the free-fermion models considered in the previous section,
let us now focus on the random Majorana chain with local interactions.

\subsection{  Majorana chain with consecutive-parity-interactions }

In this section, we focus on the case where the parity operators appearing in the Hamiltonian (Eq. \ref{HKparity})
are only those involving strings of $(2k)$ consecutive operators (instead of the general case of Eq. \ref{parityj}),
so that it is convenient to introduce the simplified notation
\begin{eqnarray}
P^{(2k)}_{[j,j+2k-1]} \equiv P^{(2k)}_{j,j+1,j+2,,..,j+2k-1}= i^k \gamma_{j} \gamma_{j+1} ... \gamma_{j+2k-2} \gamma_{j+2k-1}
\label{parityjconse}
\end{eqnarray}
The Hamiltonian of Eq. \ref{HKparity} is thus replaced by
\begin{eqnarray}
{\cal H} && = \sum_{k=1}^N  {\cal H}^{(2k)}
\nonumber \\
 {\cal H}^{(2k)} && =   \sum_{j=1}^{2N-2k+1} K^{(2k)}_{[j,j+2k-1]}  P^{(2k)}_{[j,j+2k-1]} 
\label{HKparityloc}
\end{eqnarray}
In particular, $ {\cal H}^{(2)}  $ corresponds
to the random Kitaev chain of Eq. \ref{kitaev}
\begin{eqnarray}
 {\cal H}^{(2)} && =   \sum_{j=1}^{2N-1} K^{(2)}_{[j,j+1]}  P^{(2)}_{[j,j+1]} 
= i  \sum_{j=1}^{2N-1}  K^{(2)}_{[j,j+1]}  \gamma_{j}\gamma_{j+1} 
\label{h2loc}
\end{eqnarray}
while $ {\cal H}^{(4)}  $ contains only four-Majorana-interactions between four consecutive operators
\begin{eqnarray}
 {\cal H}^{(4)} && = \sum_{j=1}^{2N-3} K^{(4)}_{[j,j+3]}  P^{(2k)}_{[j,j+3]} 
= - \sum_{j=1}^{2N-3} K^{(4)}_{[j,j+3]}\gamma_{j} \gamma_{j+1} \gamma_{j+2} \gamma_{j+3}
\label{h4loc}
\end{eqnarray}
The translation of this model in the quantum spin language is given in Eqs \ref{kitaevspins} \ref{h4locspins} \ref{HKparitylocspins} of Appendix \ref{sec_app}.

\subsection{  Renormalized consecutive parities  }

After the elimination of the two Majorana operators $(\gamma_n,\gamma_{n+1})$
corresponding to
 the biggest coupling in absolute value (Eq. \ref{k2max})
\begin{eqnarray}
\vert K^{(2)} _{n,n+1} \vert && = \max_{1 \leq j \leq 2N-1}  \vert  K^{(2)} _{j, j+1}  \vert
\label{k2maxkitaevbis}
\end{eqnarray}
the operators $\gamma_{n-1}$ and $\gamma_{n+2}$ have become neighbors.
One then needs to introduce the renormalized consecutive-parity-operators across the decimated pair 
like the one already encountered in Eq. \ref{veffreskitaev} for the Kitaev chain
\begin{eqnarray}
P^{R(2)}_{[n-1,n+2]}  \equiv i  \gamma_{n-1} \gamma_{n+2}
\label{parityRsaut}
\end{eqnarray}
Here we will need more generally the other renormalized consecutive parities 
\begin{eqnarray}
P^{R(2k-2)}_{[j,j+2k-1]}  
=  i^{k-1} \left(\gamma_j ... \gamma_{n-1} \right) \left( \gamma_{n+2}... \gamma_{n+2k-1} \right) 
\label{parityRk}
\end{eqnarray}
for $j \leq n-1$ and $j+2k-1 \geq n+2$

\subsection{ Decomposition of Eq. \ref{decomponm}  }

In the decomposition of Eq. \ref{decomponm}
\begin{eqnarray}
 {\cal H} 
&& = H_{n,n+1}^{00}  + i \gamma_{n}  H_{n,n+1}^{10}  + i \gamma_{n+1} H_{n,n+1}^{01} + i \gamma_{n}\gamma_{n+1}  H_{n,n+1}^{11}
\label{decomponmloc}
\end{eqnarray}
$H_{n,n+1}^{00} $ contains all the terms of the Hamiltonian included in $[1,..,n-1]$ or included in $[n+2,..,2N]$
\begin{eqnarray}
H_{n,n+1}^{00}  && = \sum_{k=1}^N \left(   \sum_{j=1}^{n-2k} K^{(2k)}_{[j,j+2k-1]}  P^{(2k)}_{[j,j+2k-1]} 
+  \sum_{j=n+2}^{2N-2k+1} K^{(2k)}_{[j,j+2k-1]}  P^{(2k)}_{[j,j+2k-1]} \right)
\label{Hloc00}
\end{eqnarray}
while $H_{n,n+1}^{11} $ reads in terms of the renormalized consecutive-parity-operators of Eq. \ref{parityRk}
\begin{eqnarray}
 H_{n,n+1}^{11}  && =  K^{(2)}_{[n,n+1]}
 +   \sum_{k \geq 2} 
\left(  K^{(2k)}_{[n+2-2k,n+1]} P^{(2k-2)}_{[n+2-2k,n-1]}
 + K^{(2k)}_{[n,n+2k-1]} P^{(2k-2)}_{[n+2,n+2k-1]} 
\right)
\nonumber \\
&&  +   \sum_{k \geq 2} 
 \sum_{j=n+1-2k}^{n-1} K^{(2k)}_{[j,j+2k-1]}  P^{R(2k-2)}_{[j,j+2k-1]}  
\label{Hloc11res}
\end{eqnarray}
$H_{n,n+1}^{01} $ can be obtained from all the parity operators beginning exactly at $j=n+1$,
and it is thus convenient to factor out the common operator $\gamma_{n+2}$ to rewrite
\begin{eqnarray}
 H_{n,n+1}^{01}  
&& =   \gamma_{n+2} \left(   K^{(2)}_{[n+1,n+2]}  +  \sum_{k\geq 2}   K^{(2k)}_{[n+1,n+2k]}   P^{(2k-2)}_{[n+3,n+2k]}  \right)
\label{Hloc01}
\end{eqnarray}
Similarly, $H_{n,n+1}^{10} $ can be obtained from all the parity operators ending exactly at $j+2k-1=n$,
and one can factor out the common operator $\gamma_{n-1}$ to rewrite
\begin{eqnarray}
 H_{n,n+1}^{10}  
&& = -    \left(   K^{(2)}_{[n-1,n]} + \sum_{k \geq 2}   K^{(2k)}_{[n+1-2k,n]} P^{(2k-2)}_{[n+1-2k,n-2 ]}   \right) \gamma_{n-1} 
\label{Hloc11}
\end{eqnarray}

Then their squares simplify into
\begin{eqnarray}
( H_{n,n+1}^{01} )^2 && =   \left(   K^{(2)}_{[n+1,n+2]}  +  \sum_{k\geq 2}   K^{(2k)}_{[n+1,n+2k]}   P^{(2k-2)}_{[n+3,n+2k]}  \right)^2
\nonumber \\
&& =   \sum_{k\geq 1}  ( K^{(2k)}_{[n+1,n+2k]}  )^2
+ 2  \sum_{ 1 \leq k_1 < k_2 }   K^{(2k_1)}_{[n+1,n+2k_1]} K^{(2k_2)}_{[n+1,n+2k_2]}  P^{(2k_2-2k_1)}_{[n+2k_1+1,n+2k_2]}
\label{Hloc01square}
\end{eqnarray}
and
\begin{eqnarray}
 (H_{n,n+1}^{10} )^2 && =   \left(   K^{(2)}_{[n-1,n]} + \sum_{k \geq 2}   K^{(2k)}_{[n+1-2k,n]} P^{(2k-2)}_{[n+1-2k,n-2 ]}   \right)^2
\nonumber \\
&& =   \sum_{k\geq 1} (K^{(2k)}_{[n+1-2k,n]})^2 
 +2 \sum_{1 \leq k_1 < k_2 }   K^{(2k_1)}_{[n+1-2k_1,n]} K^{(2k_2)}_{[n+1-2k_2,n]} P^{(2k_2-2k_1)}_{[n+1-2k_2,n-2k_1 ]} 
\label{Hloc10square}
\end{eqnarray}
while their commutator reads
in terms of the renormalized consecutive-parity-operators of Eq. \ref{parityRk}
 \begin{eqnarray}
&& \frac{i}{2}  [ H_{n,n+1}^{01} , H_{n,n+1}^{10} ]   
\nonumber \\
&& = 
  \left(   K^{(2)}_{[n-1,n]} + \sum_{k_1 \geq 2}   K^{(2k_1)}_{[n+1-2k_1,n]} P^{(2k_1-2)}_{[n+1-2k_1,n-2 ]}   \right) ( i \gamma_{n-1} 
 \gamma_{n+2}) \left(   K^{(2)}_{[n+1,n+2]}  +  \sum_{k_2 \geq 2}   K^{(2k_2)}_{[n+1,n+2k_2]}   P^{(2k_2-2)}_{[n+3,n+2k_2]}  \right)
\nonumber \\
&& =     K^{(2)}_{[n-1,n]}  K^{(2)}_{[n+1,n+2]}  P^{R(2)}_{[n-1,n+2]}
+      \sum_{k_1 \geq 2}   K^{(2k_1)}_{[n+1-2k_1,n]} \sum_{k_2 \geq 2}   K^{(2k_2)}_{[n+1,n+2k_2]} 
P^{R(2k_1+2 k_2-2)}_{[n+1-2k_1,n+2k_2]}  
\nonumber \\
&& + \sum_{k_1 \geq 2}   K^{(2k)}_{[n+1-2k_1,n]}  K^{(2)}_{[n+1,n+2]}  P^{R(2k_1)}_{[n+1-2k,n+2]}  
+   K^{(2)}_{[n-1,n]} \sum_{k_2 \geq 2}   K^{(2k_2)}_{[n+1,n+2k_2]}      P^{R(2k_2)}_{[n-1,n+2k_2]} 
\nonumber \\
&& =   \sum_{k_1 \geq 1} \sum_{k_2 \geq 1}   K^{(2k_1)}_{[n+1-2k_1,n]}   K^{(2k_2)}_{[n+1,n+2k_2]} 
P^{R(2k_1+2 k_2-2)}_{[n+1-2k_1,n+2k_2]}  
\label{decompa}
\end{eqnarray}

\subsection{  Renormalized Hamiltonian  }

Putting everything together, Eq \ref{veffres} yields

\begin{eqnarray}
 H^{\epsilon}_{n,n+1}  && 
= H_{n,n+1}^{00}  + \epsilon H_{n,n+1}^{11}
+ \epsilon  \frac{   (  H_{n,n+1}^{10} )^2 + (  H_{n,n+1}^{01} )^2  }{ 2 K^{(2)}_{n,n+1}   } 
+  \frac{  i  [H_{n,n+1}^{01}  , H_{n,n+1}^{10} ]  }{ 2  K^{(2)}_{n,n+1}   } 
\nonumber \\
&& =  \sum_{k=1}^N \left(   \sum_{j=1}^{n-2k} K^{(2k)}_{[j,j+2k-1]}  P^{(2k)}_{[j,j+2k-1]} 
+  \sum_{j=n+2}^{2N-2k+1} K^{(2k)}_{[j,j+2k-1]}  P^{(2k)}_{[j,j+2k-1]} \right)
\nonumber \\
&& + \epsilon  K^{(2)}_{[n,n+1]}
+ \epsilon  \sum_{k\geq 1}  \frac{  (K^{(2k)}_{[n+1-2k,n]})^2 +   ( K^{(2k)}_{[n+1,n+2k]}  )^2 }{ 2 K^{(2)}_{n,n+1}   } 
\nonumber \\
&&  + \epsilon 
  \sum_{k \geq 1} 
\left(  K^{(2k+2)}_{[n-2k,n+1]} P^{(2k)}_{[n-2k,n-1]}
+ \sum_{j=n+1-2k}^{n-1} K^{(2k+2)}_{[j,j+2k+1]}  P^{R(2k)}_{[j,j+2k+1]}  
 + K^{(2k+2)}_{[n,n+2k+1]} P^{(2k)}_{[n+2,n+2k+1]} 
\right)
\nonumber \\
&&+ \epsilon \sum_{k \geq 1 }   \sum_{ k_1 \geq 1 } 
\frac{  K^{(2k_1)}_{[n+1-2k_1,n]} K^{(2k+2k_1)}_{[n+1-2k_1-2k,n]} 
 }{  K^{(2)}_{n,n+1}   } P^{(2k)}_{[n+1-2k_1-2k,n-2k_1 ]}
\nonumber \\
&&+ \epsilon \sum_{k \geq 1 }   \sum_{ k_1 \geq 1 } 
\frac{     K^{(2k_1)}_{[n+1,n+2k_1]} K^{(2k_1+2k)}_{[n+1,n+2k_1+2k]} 
 }{  K^{(2)}_{n,n+1}   }  P^{(2k)}_{[n+2k_1+1,n+2k_1+2k]} 
\nonumber \\
&&+\sum_{k_1 \geq 1} \sum_{k_2 \geq 1}   \frac{     K^{(2k_1)}_{[n+1-2k_1,n]}   K^{(2k_2)}_{[n+1,n+2k_2]} 
   }{   K^{(2)}_{n,n+1}   } P^{R(2k_1+2 k_2-2)}_{[n+1-2k_1,n+2k_2]} 
\label{veffresconse}
\end{eqnarray}

To clarify the meaning of the various terms, it is useful to distinguish four types of contributions
\begin{eqnarray}
&& H^{\epsilon}_{n,n+1}  =  E^{\epsilon}_{n,n+1} +  H^{\epsilon Left}_{n,n+1} + H^{\epsilon Right}_{n,n+1} +  H^{\epsilon Middle}_{n,n+1} 
\label{veffresconse4terms}
\end{eqnarray}
The first term is simply the constant contribution produced directly by the decimation that depends on the energy branch $\epsilon=\pm 1$
\begin{eqnarray}
  E^{\epsilon}_{n,n+1} =  \epsilon  \left( K^{(2)}_{[n,n+1]}
+  \sum_{k\geq 1}  \frac{  (K^{(2k)}_{[n+1-2k,n]})^2 +   ( K^{(2k)}_{[n+1,n+2k]}  )^2 }{ 2 K^{(2)}_{n,n+1}   } \right)
\label{veffresconseconstant}
\end{eqnarray}
The second term contains the parity-operators localized on the left $[1,...,n-1]$ of the decimated pair
\begin{eqnarray}
  H^{\epsilon Left}_{n,n+1} 
&& =
\sum_{k \geq 1}  \sum_{l \leq n-1}
\left(  K^{(2k)}_{[l+1-2k,l]} 
+ \epsilon     K^{(2k+2)}_{[n-2k,n+1]} \delta_{l,n-1}
 + \epsilon \frac{ K^{(n-l+2k)}_{[l+1-2k,n]}   K^{(n-l)}_{[l+1,n]}  }{  K^{(2)}_{n,n+1}   }  \right) 
 P^{(2k)}_{[l+1-2k,l]} 
\label{veffresleft}
\end{eqnarray}
The third term contains the parity-operators localized on the right $[n+2,...,2N]$ of the decimated pair
\begin{eqnarray}
   H^{\epsilon Right}_{n,n+1} 
&&=  \sum_{k \geq 1}  \sum_{j \geq n+2}
\left(  K^{(2k)}_{[j,j+2k-1]}  
+ \epsilon    K^{(2k+2)}_{[n,n+2k+1]} \delta_{j,n+2}
+ \epsilon   \frac{     K^{(j-n-1)}_{[n+1,j-1]} K^{(j-n-1+2k)}_{[n+1,j+2k-1]}  }{  K^{(2)}_{n,n+1}   } 
 \right)
P^{(2k)}_{[j,j+2k-1]} 
\label{veffresright}
\end{eqnarray}
Finally the fourth term contains the renormalized parity-operators of Eq. \ref{parityRk}
that begin before the decimated pair and that end after the decimated pair
\begin{eqnarray}
  H^{\epsilon Middle}_{n,n+1} 
=  \sum_{k \geq 1} \sum_{j=n+1-2k}^{n-1} 
\left(  \epsilon  K^{(2k+2)}_{[j,j+2k+1]}  
+ \frac{     K^{(n+1-j)}_{[j,n]}   K^{(2k+j+1-n)}_{[n+1,j+2k+1]}    }{   K^{(2)}_{n,n+1}   } \right)
P^{R(2k)}_{[j,j+2k+1]}  
\label{veffresmiddle}
\end{eqnarray}

\subsection{  RSRG-X rules  }

The RSRG-X rules for the couplings between the surviving Majorana operators can be thus summarized as follows.

(i) The coupling associated to the parity operator $P^{(2k)}_{[l+1-2k,l]}  $ living on the left of the decimated pair $l \leq n-1$ (Eq. \ref{veffresleft})
follows the RG rule
\begin{eqnarray}
K^{R(2k)}_{[l+1-2k,l]} 
&&=  K^{(2k)}_{[l+1-2k,l]} 
+ \epsilon  \left(   K^{(2k+2)}_{[n-2k,n+1]} \delta_{l,n-1}
 + \frac{ K^{(2k+n-l)}_{[l+1-2k,n]}   K^{(n-l)}_{[l+1,n]}  }{  K^{(2)}_{n,n+1}   } \right)
\label{rgbefore}
\end{eqnarray}
Besides its initial value $K^{(2k)}_{[l+1-2k,l]}  $, the new contributions
comes from the 'degradation' of the higher-order couplings $K^{(2k+2)}_{[n-2k,n+1]}  $
and $K^{(2k+n-l)}_{[l+1-2k,n]}  $ of order $2k+n-l \geq 2k+2 $ and depend on the choice $\epsilon=\pm$ of the energy branch.

(ii)  The coupling associated to the parity operator   $P^{(2k)}_{[j,j+2k-1]}  $ living on the right of the decimated pair $j \geq n+2$ (Eq. \ref{veffresright})
follows the RG rule
\begin{eqnarray}
K^{R(2k)}_{[j,j+2k-1]} = K^{(2k)}_{[j,j+2k-1]}  
+ \epsilon \left(   K^{(2k+2)}_{[n,n+2k+1]} \delta_{j,n+2}
+    \frac{     K^{(j-n-1)}_{[n+1,j-1]} K^{(2k+j-n-1)}_{[n+1,j+2k-1]}  }{  K^{(2)}_{n,n+1}   } \right)
\label{rgafter}
\end{eqnarray}
Here again, besides its initial value $  K^{(2k)}_{[j,j+2k-1]}   $, the new contributions
comes from the 'degradation' of the higher-order couplings $ K^{(2k+2)}_{[n,n+2k+1]}   $
and $K^{(2k+j-n-1)}_{[n+1,j+2k-1]}  $ of order $2k+j-n-1 \geq 2k+2$ and depend on the choice $\epsilon=\pm$ of the energy branch.

(iii) The renormalized parity operator $P^{R(2k)}_{[j,j+2k+1]}   $ that begins before the decimated pair $j \leq n-1$ and 
that ends after the decimated pair $n+2 \leq j+1+2k$
(Eq \ref{veffresmiddle}) is associated to the new renormalized couplings
\begin{eqnarray}
K^{R(2k)}_{[j,j+2k+1]}  
&& = \epsilon  K^{(2k+2)}_{[j,j+2k+1]}  
+ \frac{     K^{(n+1-j)}_{[j,n]}   K^{(2k+j+1-n)}_{[n+1,j+2k+1]}    }{   K^{(2)}_{n,n+1}   }
\label{rgabove}
\end{eqnarray}
The first terms corresponds again to the  'degradation' of the higher-order coupling $K^{(2k+2)}_{[j,j+2k+1]}   $
and depends on the choice $\epsilon=\pm$ of the energy branch.
The second term is the generalization of the basic rule of Eq. \ref{kr2kitaev} concerning the Kitaev chain
and does not depend on the choice $\epsilon=\pm$ of the energy branch.
In the present procedure, this second term is the only mechanism where new higher order couplings can be generated
from two couplings of smaller orders $2k_1=n+1-j$ and $2k_2=2k+j+1-n=2k+2-2k_1 $.

In conclusion, the Majorana chain with consecutive parity-interactions of Eq. \ref{HKparityloc}
remains closed for the RSRG-X procedure with the renormalized rules described above.
To see more clearly how it works in practice, it is now useful to consider the following simplest example.

\subsection{  First RG step for the initial chain involving only two and four Majorana interactions }

Let us consider the case where the initial Hamiltonian of Eq. \ref{HKparityloc} contains only
interactions between two and four consecutive Majorana operators (Eqs \ref{h2loc} and \ref{h4loc})
\begin{eqnarray}
{\cal H}^{ini}  && = i  \sum_{j=1}^{2N-1}  K^{(2)}_{[j,j+1]}  \gamma_{j}\gamma_{j+1} 
 - \sum_{j=1}^{2N-3} K^{(4)}_{[j,j+3]}\gamma_{j} \gamma_{j+1} \gamma_{j+2} \gamma_{j+3}
\label{HKparityloc24}
\end{eqnarray}

The RSRG-X rules for the first decimation of the biggest coupling $K^{(2)}_{n,n+1} $ in absolute value are the following.

(i) The RG rule of Eq. \ref{rgbefore} for the left of the decimated pair gives new contributions only for
\begin{eqnarray}
K^{R(2)}_{[n-2,n-1]} =K^{(2)}_{[n-2,n-1]}  +  \epsilon    K^{(4)}_{[n-2,n+1]} 
\label{rgbeforea}
\end{eqnarray}
and
\begin{eqnarray}
K^{R(2)}_{[n-3,n-2]}  = K^{(2)}_{[n-3,n-2]}  + \epsilon \frac{K^{(4)}_{[n-3,n]}   K^{(2)}_{[n-1,n]}  }{  K^{(2)}_{n,n+1}   } 
\label{rgbeforeb}
\end{eqnarray}
representing the 'degradation' of the four-Majorana-couplings $K^{(4)}_{[n-2,n+1]}  $ and $K^{(4)}_{[n-3,n]} $
into contributions of couplings of order $2k=2$ that were already existing.

(ii) The RG rule of Eq. \ref{rgafter}  for the right of the decimated pair gives new contributions only for
\begin{eqnarray}
K^{R(2)}_{[n+2,n+3]} =  K^{(2)}_{[n+2,n+3]}  + \epsilon  K^{(4)}_{[n,n+3]} 
\label{rgaftera}
\end{eqnarray}
and
\begin{eqnarray}
K^{R(2)}_{[n+3,n+4]} =  K^{(2)}_{[n+3,n+4]}  + \epsilon  \frac{     K^{(2)}_{[n+1,n+2]} K^{(4)}_{[n+1,n+4]}  }{  K^{(2)}_{n,n+1}   } 
\label{rgafterb}
\end{eqnarray}
representing also the 'degradation' of the four-Majorana-couplings $K^{(4)}_{[n,n+3]}  $ and $K^{(4)}_{[n+1,n+4]} $
into contributions of couplings of order $2k=2$ that were already existing.

(iii) The RG rule of Eq. \ref{rgabove} for the renormalized parities across the decimated pair gives new couplings of various orders.
The only renormalized coupling of order $2k=2$ is
\begin{eqnarray}
K^{R(2)}_{[n-1,n+2]}  && = \epsilon   K^{(4)}_{[n-1,n+2]}  
+ \frac{     K^{(2)}_{[n-1,n]}   K^{(2)}_{[n+1,n+2]}    }{   K^{(2)}_{n,n+1}   } 
\label{rgabovea}
\end{eqnarray}
containing the 'degradation' of the four-Majorana coupling $ K^{(4)}_{[n-1,n+2]}  $ and
the renormalized contribution already seen for the Kitaev chain (Eq. \ref{kr2kitaev}).
The only renormalized couplings of order $2k=4$ are
\begin{eqnarray}
K^{R(4)}_{[n-3,n+2]}  
&& =  \frac{     K^{(4)}_{[n-3,n]}   K^{(2)}_{[n+1,n+2]}    }{   K^{(2)}_{n,n+1}   } 
\label{rgaboveb}
\end{eqnarray}
and
\begin{eqnarray}
K^{R(4)}_{[n-1,n+4]}  
&& =  \frac{     K^{(2)}_{[n-1,n]}   K^{(4)}_{[n+1,n+4]}    }{   K^{(2)}_{n,n+1}   } 
\label{rgabovec}
\end{eqnarray}
Finally, there is one new renormalized coupling of order $2k=6$
\begin{eqnarray}
K^{R(6)}_{[n-3,n+4]}  && =  \frac{     K^{(4)}_{[n-3,n]}   K^{(4)}_{[n+1,n+4]}    }{   K^{(2)}_{n,n+1}   } 
\label{rgaboved}
\end{eqnarray}

This example shows that the generation of higher-order couplings remains rather limited,
while there are many mechanisms of 'degradation' into smaller-order couplings.
So we hope that these Majorana RSRG-X procedure can be applied numerically
on large sizes without the proliferation of too many new renormalized couplings.
This numerical implementation clearly goes beyond the scope of the present work
and is left for other authors with more numerical possibilities
(see \cite{rsrgx} for the specific numerical problems related to the choice of different energy branches at each step of the RSRG-X).

\section{ Conclusion }

\label{sec_conclusion}

In this work, we have formulated a general RSRG-X procedure for random Majorana models in their Many-Body-Localized phase.
We have then derived the explicit RG rules for arbitrary quadratic Hamiltonians (free-fermions models) 
and for the random Kitaev chain with local interactions involving even numbers of consecutive Majorana fermions.
However, these two examples of application are not restrictive, and one can apply the general rule of Eq. \ref{veffres}
to any other Majorana model of interest.

Along the paper, we have stressed the advantages of the Majorana language over the usual quantum spin language
to formulate unified RG rules : 

(a) the symmetric role played by the $2N$ majorana operators allows to classify the various terms of the Hamiltonian
 by the even number $(2k)$ and the locations $1 \leq j_1<..<j_{2k}\leq 2N$ of the Majorana operators
(while the spin language requires the distinction between different types of couplings in terms of Pauli matrices as recalled in Appendix \ref{sec_app}).

(b) in the Strong Disorder Renormalization perspective, the unique elementary decimation then corresponds
 to the pairing between the two Majorana operators
that are the most strongly coupled in absolute value 
and thus leads to unified RSRG-X rules (while the spin language requires the distinction between 
the decimations of different types of couplings in terms of Pauli matrices).
In addition in free fermions models, the renormalized rule for the renormalized couplings is independent of the energy branch
(Eqs \ref{kr2kitaev} and Eq. \ref{kr2}).

(c) this 'deconstruction' into Majorana fermions suggests that the simplest Many-Body-Localized model is
actually the random Kitaev chain with interactions involving four consecutive Majorana fermions (Eq. \ref{HKparityloc24}),
while the standard model of MBL, namely the random-field XXZ chain actually corresponds to a Majorana {\it ladder }
with some degeneracy in the couplings $J^x=J^y=J^z$ (see Appendix \ref{sec_app}) so that the RSRG-X rules 
are more complicated as described in Ref \cite{c_rsrgt}.
It would be thus interesting in the future to apply numerically the RSRG-X rules to the Simplest MBL model of Eq. \ref{HKparityloc24} as discussed after Eq. \ref{rgaboved}.

\appendix

\section{ Dictionary between Majorana fermions, Dirac fermions and quantum spin chains } 

\label{sec_app}

\subsection{ Majorana formulation of Dirac Fermions models }

In any dimension, a model involving $N$ Dirac Fermions described by annihilation and creation operators $(c_j,c_j^{\dagger})$ for $j=1,..,N$
satisfying the canonical anti-commutation relations
\begin{eqnarray}
\{c_j,c_k \} &&= 0 =\{c_j^{\dagger},c_k^{\dagger} \} 
\nonumber \\
\{c_j,c_k^{\dagger} \}  && = \delta_{jk}
\label{diracanti}
\end{eqnarray}
can be rewritten in terms of the real and imaginary parts 
\begin{eqnarray}
\gamma_{2j-1} && \equiv    c_j^{\dagger} + c_j 
\nonumber \\
\gamma_{2j}  && \equiv   i (c_j^{\dagger} - c_j) 
\label{diracmaj}
\end{eqnarray}
that correspond to the $(2N)$ Majorana operators of Eqs \ref{hermi} , \ref{squareunity} , \ref{anticomm} by the simple substitution
\begin{eqnarray}
c_j && = \frac{\gamma_{2j-1} + i \gamma_{2j} }{2}
\nonumber \\
 c_j^{\dagger} && = \frac{\gamma_{2j-1} - i \gamma_{2j} }{2}
\label{cdiracmaj}
\end{eqnarray}

\subsection{ Majorana formulation of quantum spin chains   }

For a chain of $N$ quantum spins described by Pauli matrices, 
if the Hamiltonian commutes with the total parity
\begin{eqnarray}
P^{tot}= \prod_{k=1}^N (-\sigma_j^z)
\label{sigmaz}
\end{eqnarray}
it can be rewritten via the standard Jordan-Wigner transformation in terms of the $(2N)$ string operators 
\begin{eqnarray}
\gamma_{2j-1} && \equiv \left( \prod_{k=1}^{j-1} \sigma_k^z \right) \sigma_j^x
\nonumber \\
\gamma_{2j} && \equiv  \left( \prod_{k=1}^{j-1} \sigma_k^z \right)  \sigma_j^y 
\label{sigmaxy}
\end{eqnarray}
that correspond to the $(2N)$ Majorana operators of Eqs \ref{hermi} , \ref{squareunity} , \ref{anticomm}.

For instance, the simplest local terms commuting with the total parity have for translation
\begin{eqnarray}
 \sigma^z_j && = -i  \gamma_{2j-1} \gamma_{2j} 
\nonumber \\
\sigma_j^x \sigma^x_{j+1} && = -i    \gamma_{2j} \gamma_{2j+1}
\nonumber \\
\sigma_j^y \sigma^y_{j+1}  && = i  \gamma_{2j-1}  \gamma_{2j+2} 
\nonumber \\
\sigma_j^z \sigma^z_{j+1} && = -\gamma_{2j-1} \gamma_{2j} \gamma_{2j+1} \gamma_{2j+2} 
\nonumber \\
\sigma_j^x \sigma^x_{j+2} && = - \gamma_{2j} \gamma_{2j+1} \gamma_{2j+2} \gamma_{2j+3}
\label{dico}
\end{eqnarray}

In particular, the random transverse field Ising chain (RTFIC) translates into the random Kitaev chain of Eq. \ref{kitaev} 
 \begin{eqnarray}
H^{RTFIC} && =  - \sum_{j=1}^N h_j \sigma_j^z - \sum_{j=1}^{N-1} J_j^x \sigma_j^x \sigma_{j+1}^x 
\nonumber \\
&& = i \sum_{j=1}^{2N-1} K^{(2)}_{j,j+1} \gamma_j \gamma_{j+1}  =H^{Kitaev} 
\label{kitaevspins}
\end{eqnarray}
with the correspondence
\begin{eqnarray}
h_j &&= K^{(2)}_{2j-1,2j}
\nonumber \\
 J_j^x && =K^{(2)}_{2j,2j+1}
\label{hJK}
\end{eqnarray}
The well-known duality between fields $h_j$ and couplings $J_j^x$ thus becomes obvious in the Majorana language where
 they correspond to odd and even two-Majorana-couplings respectively.

The additional interactions between four consecutive Majorana operators of the Hamiltonian
 $ {\cal H}^{(4)}  $ of Eq. \ref{h4loc} translates into
\begin{eqnarray}
 {\cal H}^{(4)} &&  
= - \sum_{j=1}^{2N-3} K^{(4)}_{[j,j+3]}\gamma_{j} \gamma_{j+1} \gamma_{j+2} \gamma_{j+3}
\nonumber \\
&& =  \sum_{j=1}^{N-1} K^{(4)}_{[2j-1,2j+2]} \sigma_j^z \sigma^z_{j+1} 
 + \sum_{j=1}^{N-2} K^{(4)}_{[2j,2j+3]} \sigma_j^x \sigma^x_{j+2}
\label{h4locspins}
\end{eqnarray}
The first term in $\sigma_j^z \sigma^z_{j+1}  $ is the standard nearest-neighbor interaction term in the field of quantum spin chains,
while the second term $ \sigma_j^x \sigma^x_{j+2}$ between next-nearest-neighbor is less usual but nevertheless interesting to consider,
as discussed in \cite{affleck,grover} for the case of pure Majorana models.

More generally, the Hamiltonian ${\cal H}^{(2k)}  $ of Eq. \ref{HKparityloc} involving the consecutive parity operators of Eq. \ref{parityjconse}
reads in the spin language
\begin{eqnarray}
 {\cal H}^{(2k)} && =   \sum_{j=1}^{2N-2k+1} K^{(2k)}_{[j,j+2k-1]}   i^k \gamma_{j} \gamma_{j+1} ... \gamma_{j+2k-2} \gamma_{j+2k-1}
\nonumber \\
&& = (-1)^k \sum_{j=1}^{N+1-k} K^{(2k)}_{[2j-1,2j+2k-2]}  \sigma_j^z \sigma_{j+1}^z \sigma_{j+2}^z... \sigma^z_{j+k-1} 
 +  (-1)^k\sum_{j=1}^{N-k} K^{(2k)}_{[2j,2j+2k-1]} \sigma_j^x \sigma^x_{j+k}
\label{HKparitylocspins}
\end{eqnarray}
where the first term involves $k$ consecutive Pauli matrices $\sigma^z$, while the second term involves only two Pauli matrices $\sigma^x$ separated by the distance $k$.

As a final remark, let us mention that the Jordan Wigner transformation of Eq. \ref{sigmaxy} is of course specific to one dimension,
but for certain bidimensional quantum spin models, other relations have been introduced between quantum spins and Majorana fermions
\cite{kitaev_honey,feng,nussinov}.

\end{document}